\documentclass[twocolumn]{revtex4}
\usepackage{graphicx}

\begin{document}
\title{The Modified Schrodinger Poisson Equation --- Quantum Polytropes}
\author{Jeremy Heyl}
\email{heyl@phas.ubc.ca}
\author{Matthew W. Choptuik}
\author{David Shinkaruk}
\affiliation{Department of Physics and Astronomy,
University of British Columbia, Vancouver BC V6T 1Z1 Canada}
\begin{abstract}
Axions and axion-like particles are a leading model for the dark matter in the Universe; therefore, dark matter halos may be boson stars in the process of collapsing.  We examine a class of static boson stars with a non-minimal coupling to gravity.  We modify the gravitational density of the boson field to be proportional to an arbitrary power of the modulus of the field, introducing a non-standard coupling.  We find a class of solutions very similar to Newtonian polytropic stars that we denote ``quantum polytropes.'' These quantum polytropes are supported by a non-local quantum pressure and follow an equation very similar to the Lane-Emden equation for classical polytropes.  Furthermore, we derive a simple condition on the exponent of the non-linear gravitational coupling, $\alpha>8/3$, beyond which the equilibrium solutions are unstable.  
\end{abstract}

\maketitle
\section{Introduction}

Bosonic dark matter possibly in the form of low-mass axions is a leading contender to explain some inconsistencies in the standard cold dark matter model (CDM) \cite{2016arXiv161008297H}.  It is inspired from both a theoretical point of view \cite{2016arXiv160705769A} as emerging from string theory and observationally where bosonic dark matter can address some potential discrepancies in the standard CDM model \cite{2015PhRvD..92j3510B,2016PhRvD..93j3533K,2016PDU....12...50P}.  Because the bosons can collapse to form a star-like object \cite{2016JCAP...07..009V,2016PhRvD..94d3513S}, small-scale structure would be different if the dark matter were dominated by light bosons.  Furthermore the collisions of these dark matter cores or boson stars would result in potentially observable interference \cite{2011PhRvD..83j3513G}.  It is these boson stars that are the focus of this investigation. 

The Schrodinger-Poisson equation provides a model for a boson star \cite{PhysRev.187.1767} in the Newtonian limit.  We will explore the solutions to the Schrodinger-Poisson equation with a small yet non-trivial modification. The modified Schrodinger-Poisson equation is given by the following two equations
\begin{equation}
  i \frac{\partial \psi}{\partial t} = -\frac{1}{2} \nabla^2 \psi + V \psi
  \label{eq:1}
\end{equation}
where
\begin{equation}
  \nabla^2 V = |\psi|^\alpha
  \label{eq:2}
\end{equation}
where we have taken $m=1$ and $4\pi G=1$.  For $\alpha=2$ this equation is the well-known non-relativistic limit of the Klein-Gordon equation coupled to gravity \cite{2012CQGra..29u5010G}.  
For $\alpha\neq 2$, this is not the case.  Although the Newtonian limit of a self-gravitating scalar field
with a potential of the form $|\psi|^\alpha$ would yield
Eq.~\ref{eq:2}, one would not get Eq.~\ref{eq:1}, the Schrodinger equation, as the non-relativistic limit for the dynamics of the scalar field.   Instead Eqs.~\ref{eq:1} and~\ref{eq:2} result as the Newtonian limit of a relativistic scalar field with a non-minimal coupling to gravity such as the following scalar-tensor action
\begin{equation}
S = \int d^4 x \sqrt{-g} \left [ \frac{R+L_m}{|\psi|^{\alpha-2}} +  \partial^\mu \bar \psi \partial_\mu \psi -  |\psi|^2 \right]
\label{eq:64}
\end{equation}
where $R$ is the Ricci scalar, $g$ is the determinant of the metric and $L_m$ is the Lagrangian density of the matter.

The small change in Eq.~\ref{eq:2} yields a new richness to the solutions for Newtonian boson stars that we will call ``quantum polytropes'' for reasons that will become obvious later.  Although authors have considered other modifications to the Schrodinger-Poisson equation such as an electromagnetic field \cite{2015GReGr..47....1M} or non-linear gravitational terms \cite{2015CQGra..32f5010F,2016CQGra..33g5002F}, the non-linear coupling of the gravitational source proposed here is novel.

\section{Homology}
\label{sec:homology}

We can examine how the equations change under a homology or scale transformation.  Let us replace the four variables with scaled versions as
\begin{equation}
  \psi \rightarrow A \psi,
  V \rightarrow A^a V,
  r \rightarrow A^b r ~\textrm{and}~
  t \rightarrow A^c t
  \label{eq:3}
\end{equation}
and try to find the values of the exponents that result in the same equations again.
\begin{equation}
  i A^{1-c} \frac{\partial \psi}{\partial t} = -\frac{1}{2} A^{1-2b} \nabla^2 \psi + A^{1+a} V \psi
  \label{eq:4}
\end{equation}
and
\begin{equation}
  A^{a-2b}  \nabla^2 V = A^{\alpha} |\psi|^\alpha.
  \label{eq:5}
\end{equation}
This yields the following equations for the exponents
\begin{equation}
  1-c=1-2b=1+a, a-2b=\alpha
  \label{eq:6}
\end{equation}
and the following scalings
\begin{equation}
  \psi \rightarrow A \psi,   V \rightarrow  A^{\alpha/2} V, r \rightarrow A^{-\alpha/4} r ~\textrm{and}~ t \rightarrow A^{-\alpha/2} t.
  \label{eq:7}
\end{equation}
The total norm of a solution which is conserved is given by
\begin{equation}
  N = \int_0^\infty 4\pi r^2 |\psi|^2 d r
  \label{eq:8}
\end{equation}
and scales under the homology transformation as $N \rightarrow A^{(8-3\alpha)/4}$.  For a static solution the value of the energy eigenvalue ($E$) scales as $A^{\alpha/2}$.  Because the solution is not normalized, the total energy will scale as the product of the eigenvalue and the norm, yielding $A^{(8-\alpha)/4}$.

We see that for $\alpha=8/3$, one can increase the central value of the wavefunction $\psi(0)$ without changing the norm but increasing the magnitude of the energy resulting in a more bound configuration.  For larger values of $\alpha$ the value of the norm decreases.  We can
argue that the this decrease in the norm results in an unstable configuration.  Let us divide the configuration arbitrarily into a central region and an arbitrarily small envelope.  If we let the central region collapse slightly, energy is released but according to the decrease in norm of this central region, we still have some material left to add to the diffuse envelope to carry the excess
energy and the process can continue to release energy.  The star is unstable. For $\alpha<8/3$ the slight collapse results in an increase in the norm of the central region but there is no material to add except from the arbitrarily small envelope, so the collapse fails.  If we let the star expand a bit in this case, the norm decreases.  However, the expansion costs energy so the star is again stable to the radial perturbation. 

For $\alpha=8/3$ the norm is independent of $\psi(0)$ and only depends on the number of nodes of the solution; therefore, it is natural to compare solutions for different values of $\alpha$ by choosing to normalize them to the value of the norm for $\alpha=8/3$ for the corresponding state. 

\section{Real Equations of Motion}
\label{sec:real-equat-moti}

We would like examine the static solutions of
Eq.~(\ref{eq:1}) and~Eq.~(\ref{eq:2}).  We will make the following substitution
\begin{equation}
  \psi = a e^{iS}
  \label{eq:9}
\end{equation}
where the functions $a=a({\bf r},t)$ and $S=S({\bf r},t)$ are explicitly real.  This results in the three equations
\begin{eqnarray}
  \frac{\partial a^2}{\partial t} + \nabla \cdot \left ( a^2 \nabla S \right ) &=& 0, \label{eq:10}\\
  \frac{\partial S}{\partial t} + \frac{1}{2} \left ( \nabla S \right )^2 + V - \frac{1}{2 a} \nabla^2 a &=& 0, \label{eq:11}\\
  \nabla^2 V &=& |a|^\alpha\label{eq:12}
\end{eqnarray}
that in analogy with fluid mechanics we can call the continuity
equation, the Euler equation and the Poisson equation.   We can develop
this analogy further by defining $\rho=a^2$ and ${\bf U}=\nabla S$ and taking
the gradient of Eq.~(\ref{eq:11}) to yield
\begin{eqnarray}
  \frac{\partial \rho}{\partial t} + \nabla \cdot \left ( \rho {\bf U} \right ) &=& 0, \label{eq:13}\\
  \frac{\partial {\bf U}}{\partial t} + \left ( {\bf U} \cdot \nabla \right ) {\bf U} + \nabla \left ( V -  \frac{1}{2 a} \nabla^2 a \right ) &=& 0. \label{eq:14}
\end{eqnarray}
These are simply the Madelung equations \cite{1927ZPhy...40..322M}. If we had retained constants such as the Planck constant $h$ in the Schrodinger equation, we would find the that final term in the Euler equation is proportional to $h^2$ and is a quantum mechanical specific enthalpy,
\begin{equation}
  w = - \frac{1}{2 a} \nabla^2 a.
  \label{eq:15}
\end{equation}
Furthermore, because ${\bf U}=\nabla S$ the vorticity of the flow must vanish.

We can exploit the fluid analogy further to write the equations in a
Lagrangian form using
\begin{equation}
  \frac{d}{dt} = \frac{\partial}{\partial t} + \left ({\bf U} \cdot \nabla\right )
  \label{eq:16}
\end{equation}
to yield
\begin{eqnarray}
  \frac{d \rho}{d t} + \rho \nabla \cdot {\bf U} &=& 0, \label{eq:17}\\
  \frac{d {\bf U}}{d t} + \nabla \left ( V -  \frac{1}{2 a} \nabla^2 a \right ) &=& 0.\label{eq:18}
\end{eqnarray}
A static solution to these equations will have $S=-E t$ in analogy with the time-independent Schrodinger equation and $a=a({\bf r})$
where $a$ satisfies
\begin{equation}
  -E a -\frac{1}{2} \nabla^2 a + V a = 0.\label{eq:19}
\end{equation}
An alternative treatment would exploit the fact that ${\bf U}$ must vanish for this static solution so
\begin{equation}
 \frac{1}{2a} \nabla^2 a = V + \textrm{constant}
  \label{eq:20}
\end{equation}
where we can identify the constant with the value of $E$ in Eq.~\ref{eq:19}.
Furthermore we have
\begin{equation}
  \nabla^2 V = |a|^\alpha = \nabla^2 \left ( \frac{1}{2a} \nabla^2 a \right )
  \label{eq:21}
\end{equation}
so if we specialize to a spherically symmetric solution, we have
\begin{equation}
- \frac{1}{r} \frac{d^2}{dr^2} \left [ \frac{1}{2a} \frac{d^2}{dr^2} \left ( r a \right ) \right ]  + |a|^\alpha = 0
  \label{eq:22}
\end{equation}
This equation is reminiscent of the Lane-Emden equation for polytropes 
\begin{equation}
  \frac{1}{r}  \frac{d^2}{d r^2} \left ( r \theta \right )   + \theta^n = 0,
  \label{eq:23}
\end{equation}
so a natural designation for these objects is ``quantum polytropes.''

Our equation is of course fourth order with a negative sign.  We must
supply four boundary conditions.  In principle these are 
\begin{eqnarray}
  a(0) &=& a_0,  \\
  \left .\frac{da}{dr}\right|_{r=0}&=&0, \\
  \left . -\frac{1}{2a r} \frac{d^2 (ra)}{dr^2} \right |_{r=0} &=& w_0
  \label{eq:24}
\end{eqnarray}  
and  
\begin{equation}
  \frac{d}{dr} \left [ \frac{1}{2a r} \frac{d^2 (ra)}{dr^2} \right ]_{r=0} = 0.
  \label{eq:25}
\end{equation}  
Of course not all values of $a_0$ and $w_0$ will yield physically reasonable configurations, so we must vary $w_0$ for example to find solutions such that $\lim_{r\rightarrow \infty} a(r) = 0$.  However, using the scaling rules in \S~\ref{sec:homology}, once the value of $w_0$ is determined, one can rescale the solution.

In the case of the Lane-Emden equation for $n>5$ one can find solutions where $\theta=0$ at a finite radius, {\em i.e.} a star with a surface.  From Eq.~\ref{eq:19} we find that
\begin{equation}
  E = -\lim_{r\rightarrow\infty} \frac{1}{2a r} \frac{d^2
    (ra)}{dr^2} = \lim_{r\rightarrow\infty} w(r).
  \label{eq:26}
\end{equation}
Therefore, if $E\neq 0$, the quantum system must extend to an
infinite radius.

To examine the regularity conditions near the centre, let us expand
the solution near the centre as
\begin{equation}
  a(r) = a_0 + a_2 r^2 + a_4 r^4
  \label{eq:27}
\end{equation}
where we have dropped the odd terms to ensure that the derivative of
the density and the derivative of the enthalpy vanish at the centre.
We find that
\begin{equation}
  w_0 = -3 \frac{a_2}{a_0}
  \label{eq:28}
\end{equation}
and
\begin{equation}
  a_4 = \frac{a_0^\alpha a_0^2 + 18 a_2^2}{60 a_0} = a_0 \left ( \frac{|a_0|^\alpha}{60} + \frac{w_0^2}{30} \right ).
  \label{eq:29}
\end{equation}

\begin{figure}
  \includegraphics[width=\columnwidth]{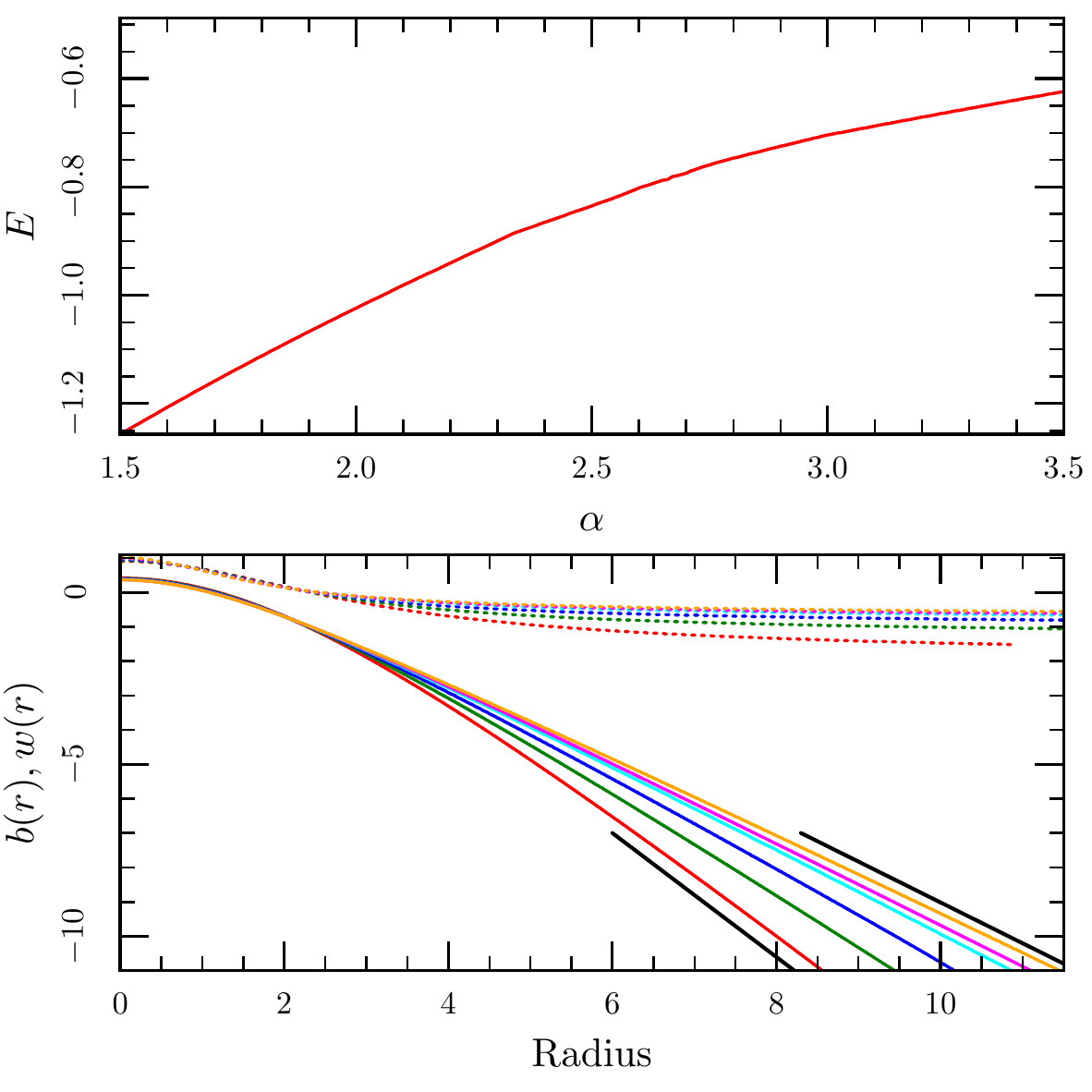}
  \caption{ Upper: The energy eigenvalue of ground state. As discussed in the text, we choose to normalize the ground states states to have the same normalization of the $\alpha=8/3$ ground-state
    solution.  Lower: The solid curves trace ground state
   The function is given by the equation $b(r)=\ln
    \psi(r)$. The solutions from bottom to top are
    $\alpha=1,1.5,2,2.5,8/3$ and 3.  The black lines show the expected
    slope of the solution for large values of $r$ from
    Eq.~(\ref{eq:34}) for $\alpha=1$ and 3. The dotted curves give the value of $w(r)$
    for the same states from bottom to top.}
  \label{fig:ground_states}
\end{figure}

As we would like to focus on the ground state where the function $a(r)$ has no nodes, we can also make the substitution that $a(r)=e^b$
which yields a simpler differential equation for $b(r)$,
\begin{equation}
b^{(4)}(r) = 2 \left [ e^{\alpha b} - \frac{2}{r} \left ( b' b''+ b''' \right ) - b' b''' -
    \left (b'' \right)^2 \right ]
  \label{eq:30}
\end{equation}
and
\begin{equation}
  w = -\frac{b''+\left(b'\right)^2}{2}-\frac{b'}{r}.
  \label{eq:31}
\end{equation}
An examination of Eq.~\ref{eq:30} and~\ref{eq:31} yields the boundary conditions at $r=0$,
\begin{eqnarray}
  b'(0)&=&0,\\
  b''(0)&=&-\frac{2}{3} w_0,\\
  b'''(0)&=&0
  \label{eq:32}
\end{eqnarray}
so a series expansion about $r=0$ for $b(r)$ yields
\begin{equation}
  b(r) = b_0 - \frac{w_0}{3} r^2 + \frac{3 e^{\alpha b_0} - 4 w_0^2}{180} r^4 + {\cal O}(r^5)
  \label{eq:33}
\end{equation}
Furthermore, we can examine the behavior at large distances from Eq.~\ref{eq:26} to
find that
\begin{equation}
  \lim_{r\rightarrow\infty} b(r) \approx -r \sqrt{-2E} = -r\sqrt{-2w}
  \label{eq:34}
\end{equation}
Fig.~\ref{fig:ground_states} depicts the ground state wavefunction
$b(r)=\ln \psi(r)$ for various values of $\alpha$.  The wavefunction
is normalized such that $N = \int dV |\psi|^2$ is constant.
Furthermore, we have verified that the scaling relations of
\S~\ref{sec:homology} hold for these solutions.  At fixed total
normalization the wavefunction is more spatial extended as $\alpha$
increases.  The slope for large values of $r$ decreases gradually with
increasing $\alpha$ reflecting the modest decrease in the binding
energy as $\alpha$ increases.

\section{Excited States}
\label{sec:excited-states}

To study the excited states \cite{1998CQGra..15.2733M} where $a(r)$ may have nodes, we have a more complicated differential equation of the form
\begin{equation}
a^{(4)}(r) = 2 a |a|^\alpha - \frac{4 a'''}{r} + \frac{N_1}{a} + \frac{N_2}{a^2}
  \label{eq:65}
\end{equation}
where
\begin{equation}
  N_1 = 2 a' a''' + \left (a''\right)^2 + \frac{8}{r} a' a''
  \label{eq:66}
\end{equation}
and
\begin{equation}
  N_2 = -2 \left (a'\right)^2 a'' - \frac{4}{r} \left (a'\right)^3.
  \label{eq:67}
\end{equation}
Rather than deal with these singular points we can return to the
coupled differential equations~\ref{eq:1} and~\ref{eq:2} to examine
the excited states.

We will make the substitutions that $u=\psi(r) r e^{-iEt}$ and
$v=V(r) r$ to yield the following equations
\begin{equation}
  E u = -\frac{1}{2}  u'' + \frac{v u}{r}
  \label{eq:68}
\end{equation}
and
\begin{equation}
  v'' = |u|^\alpha r^{1-\alpha},
  \label{eq:69}
\end{equation}
where we have focused on spherically symmetric configurations.
Because equations~\ref{eq:1} and~\ref{eq:2} are non-linear we cannot
follow the strategy of expanding the solutions in terms of spherical
harmonics to yield a simple solution beyond spherical symmetry.  The
general solution is beyond the scope of this paper.

\begin{figure*}
  \includegraphics[width=\columnwidth]{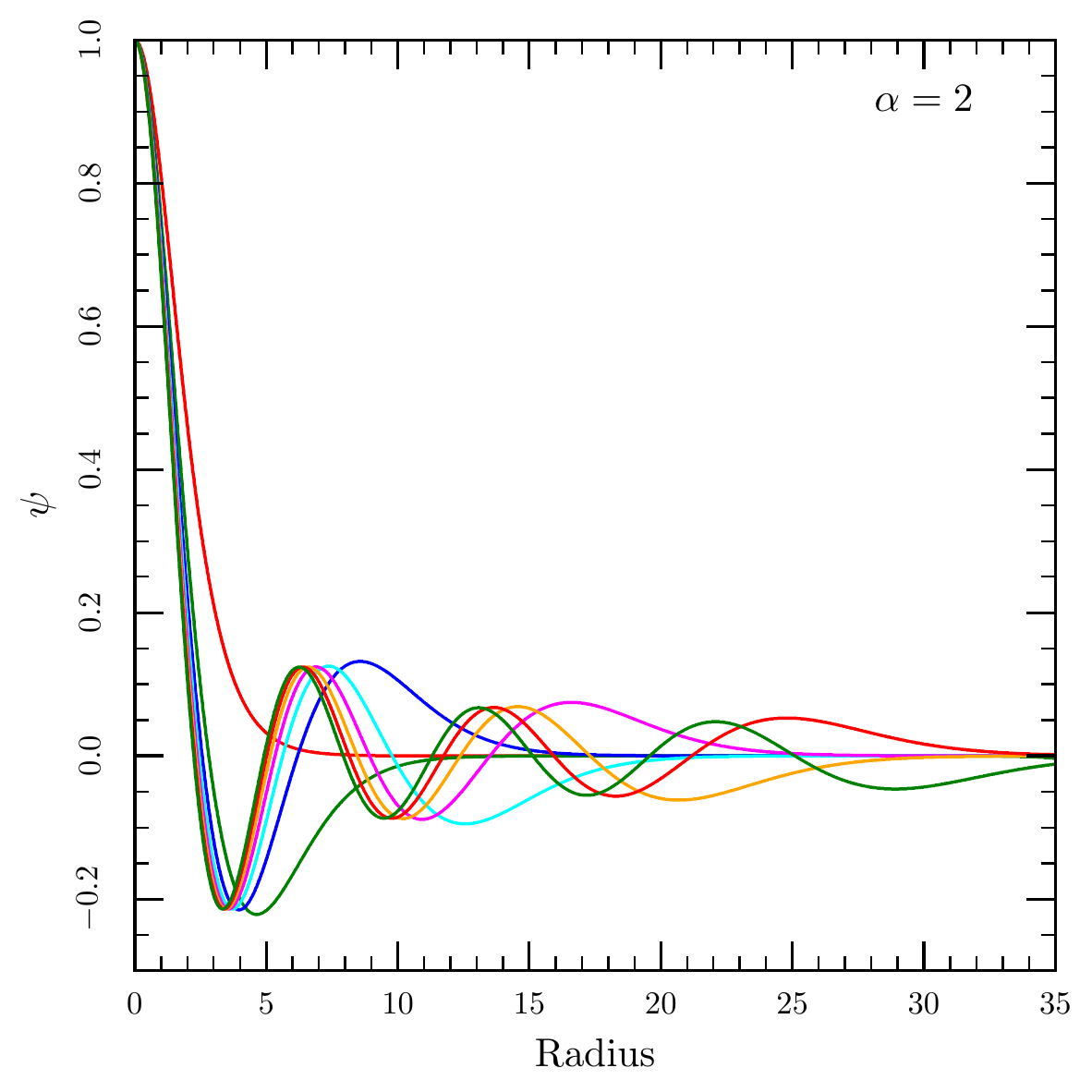}
  \includegraphics[width=\columnwidth]{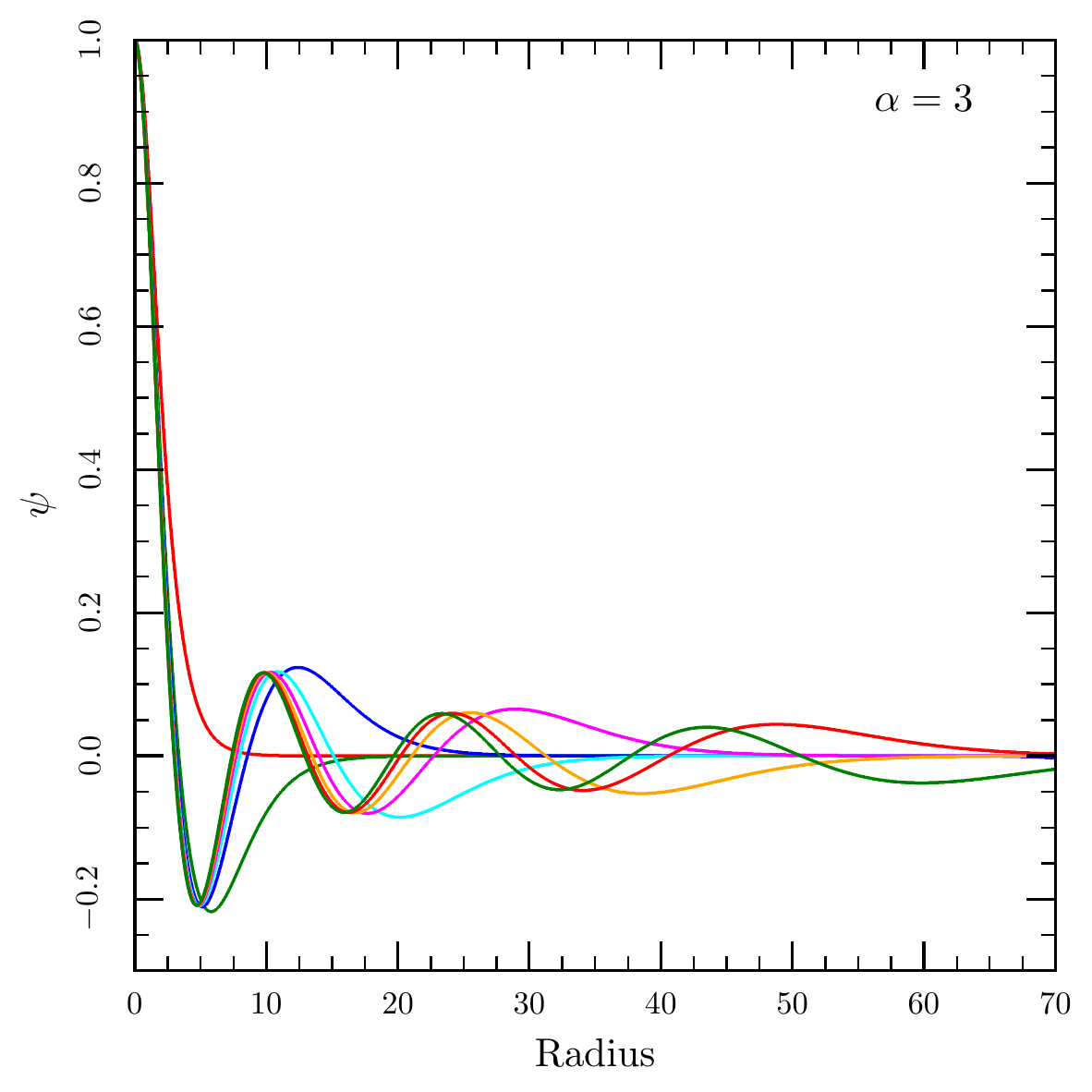}
  \caption{The ground and first seven excited states for $\alpha=2$ and $\alpha=3$ where the wavefunction is normalized such that $\psi(0)=1$.}
  \label{fig:psi23}
\end{figure*}

We must supply four boundary conditions for the functions $u$ and $v$
and these are $u=0$, $u'=\psi(0)$, $v=0$ and $v'=V(0)$ where we take
$V(0)=0$ because we can shift both the value of $E$ and $V(r)$ by a
constant and retain the same equations.  We generally shift $E$ and
$V(r)$ such that $\lim_{r\rightarrow\infty} V(r)=0$. We can also take
$\psi(0)=1$ and scale the resulting solution using the scaling
relations in \S~\ref{sec:homology}.  Finally only specific values of
$E$ will result in normalizable solutions, so we shoot from the origin to
large radii and find the values of $E$ that result in normalizable
solutions.  Fig.~\ref{fig:psi23} depicts the ground state and the
excited states for $\alpha=2$ and $\alpha=3$ where the wavefunction
has been normalized such that $\psi(0)=1$.  It is important to note
that the various states correspond to different total normalizations,
{\em i.e.} different numbers of particles.  Furthermore, we will call the
ground state the state without any nodes and excited states states
with nodes, so the quantum number $n$ denotes the number of anti-nodes
or extrema, starting with one; therefore, Fig.~\ref{fig:psi23} shows
the wavefunctions for $n=1$ to $n=8$.  The wavefunctions for
$\alpha=2$ and $\alpha=3$ appear quite similar modulo a size scaling.
The $\alpha=3$ wavefunctions with this particular normalization extend
over a larger range in radius than the $\alpha=2$ wavefunctions.

\begin{figure}
    \includegraphics[width=\columnwidth]{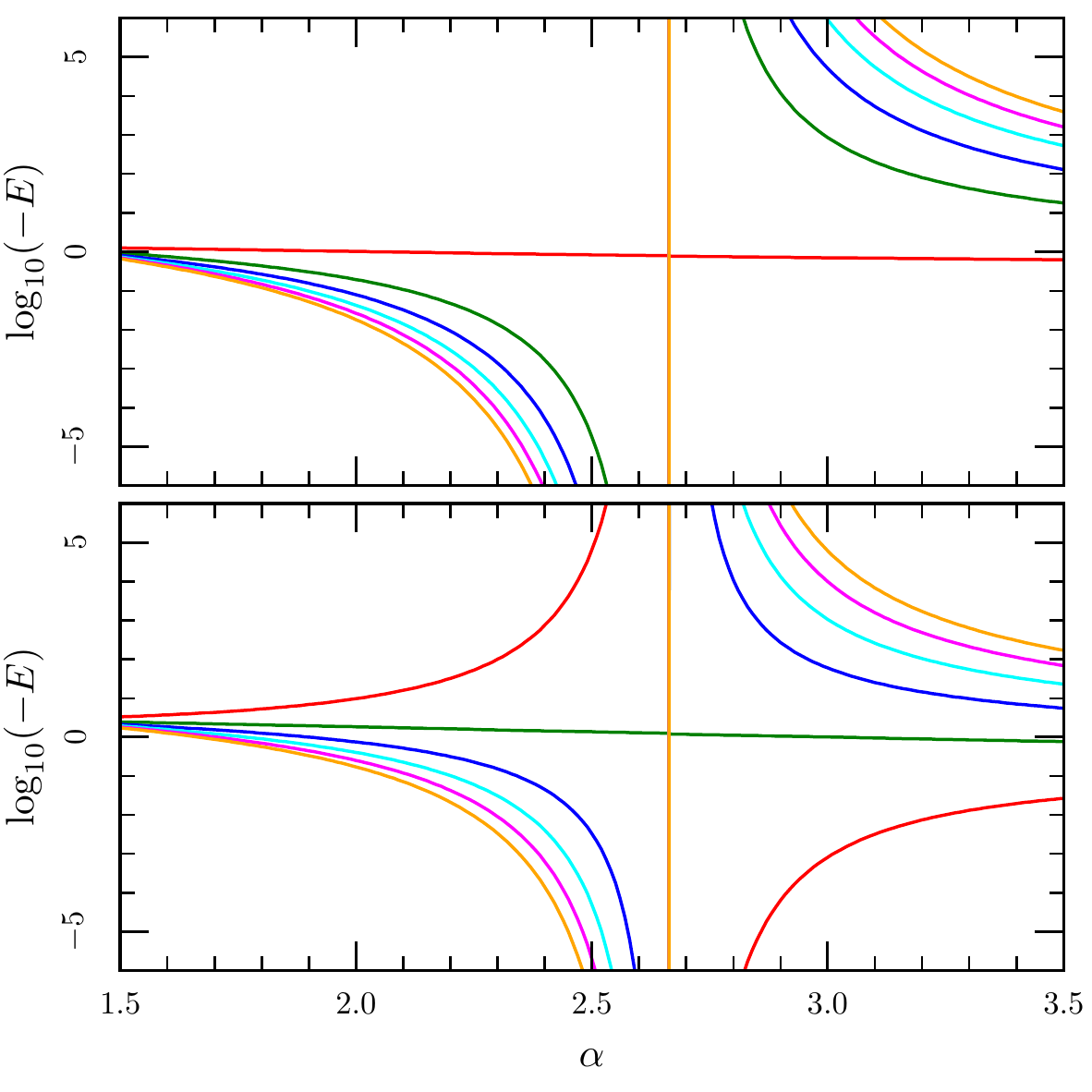}
  \caption{Energy eigenvalue of the states for a number of particles
    fixed to that of the ground state of the $\alpha=8/3$
    configuration (upper panel) and to the first excited state (lower
    panel). In both cases more bound states lie at the top.  On the
    left-hand side ($\alpha<8/3$) of both plots the states from top to bottom are
    $n=1, 2, 3, 4, 5$ and 6.  On the right-hand side ($\alpha>8/3$) the ordering is
    reversed, {\em i.e.} from top to bottom the states are $n=6,5,4,3,2$ and 1.}
  \label{fig:ominter}
\end{figure}

Of course, what is most interesting are the configurations for a fixed
number of particles, so a particular value of $N=\int dV |\psi|^2$.
For $\alpha\neq 8/3$ the total normalization, $N$, can take any value.
However, for $\alpha=8/3$ the normalization is fixed to the values of the ground and
the various excited states.  Fig.~\ref{fig:ominter} depicts the
binding energy as a function of $\alpha$ for two particular choices of
normalization.  As both the logarithm of the normalization and the
value of the energy $E$ are smooth functions of $\alpha$ for
$\psi(0)=1$, we calculate these values for $\alpha=2,7/3,8/3,3$ and
$10/3$ and interpolate or extrapolate over the plotted range.  We then
use the scaling relations from \S~\ref{sec:homology} to find the
eignenvalues for a particular normalization.  

What is most striking about the energy levels is that for $\alpha<8/3$
we have the normal ordering where states with more nodes are less
bound.  For $\alpha>8/3$ as the number of nodes increases so does the
binding energy of the state.  The energy levels are not bounded from
below in this case, a hallmark of instability.  For the limiting case
$\alpha=8/3$ we see that at most one state is bound for a particular
total normalization, $N$, but that its energy is arbitrary because we can
scale the value of the wavefunction which changes the energy eigenvalue without
changing the total normalization.

\section{Perturbations}
\label{sec:perturbations}

The results from scaling in \S~\ref{sec:homology} and from the
examination of the excited states in \S~\ref{sec:excited-states} give
very strong hints that quantum polytropes with $\alpha>8/3$ are
unstable.  We will prove that $\alpha>8/3$ is a sufficient condition
for instability for an arbitrary stationary configuration. Let us take a constant background and examine small perturbations of the form
\begin{equation}
  a = a_0 + a_1({\bf r},t)~\textrm{and}~{\bf U} = {\bf U}_1({\bf r},t)\label{eq:35}
\end{equation}
so we have
\begin{eqnarray}
2 a_0  \frac{\partial a_1}{\partial t} + a_0^2 \nabla \cdot {\bf U}_1 &=& 0,\label{eq:36} \\
  \frac{\partial {\bf U_1}}{\partial t} + \nabla \left ( V_1 -  \frac{1}{2 a_0} \nabla^2 a_1 \right ) &=& 0. \label{eq:37}
\end{eqnarray}
Now if we take the time derivative of Eq.~(\ref{eq:36}) and the divergence
of Eq.~(\ref{eq:37}), we can combine the equations to yield
\begin{equation}
  2 a_0 \frac{\partial^2 a_1}{\partial t^2} - a_0^2 \nabla^2 V_1 + \frac{a_0}{2} \nabla^4 a_1 = 0
  \label{eq:38}
\end{equation}
and
\begin{equation}
   \frac{\partial^2 a_1}{\partial t^2} - \frac{\alpha}{2} a_1 |a_0|^\alpha + \frac{1}{4} \nabla^4 a_1 = 0.
  \label{eq:39}
\end{equation}
If we expand the perturbations in Fourier components we obtain the following dispersion relation
\begin{equation}
\omega^2 =  \frac{k^4}{4} - \frac{\alpha}{2} |a_0|^\alpha 
  \label{eq:40}
\end{equation}
where the first term is the standard result for the deBroglie
wavelength of a particle and the second term is due to the
self-gravity of the perturbation.

We can be a bit more sophisticated now and assume
that small perturbations lie near a static solution so
\begin{equation}
  a = a_0({\bf r}) + a_1({\bf r},t)~\textrm{and}~{\bf U} = {\bf U}_1({\bf r},t)\label{eq:41}
\end{equation}
thus we have
\begin{eqnarray}
  2 a_0  \frac{\partial a_1}{\partial t} + \nabla \cdot \left ( a_0^2 {\bf U}_1 \right )  &=& 0,\label{eq:42}\\
  \frac{\partial {\bf U_1}}{\partial t} + \nabla \left ( \frac{a_1}{a_0} V_0 + V_1 -  \frac{1}{2 a_0} \nabla^2 a_1 \right ) &=& 0. \label{eq:43}
\end{eqnarray}
and if we take the time derivative of Eq.~(\ref{eq:42}), we can
combine the equations to yield
\begin{equation}
  2 a_0 \frac{\partial^2 a_1}{\partial t^2} =
  \nabla \cdot \left [ a_0^2
     \nabla \left ( \frac{a_1}{a_0} V_0 + V_1 -  \frac{1}{2 a_0} \nabla^2 a_1 \right ) \right ].
\label{eq:44}
\end{equation}
Furthermore, the perturbation of the potential satisfies
\begin{equation}
  \nabla^2 V_1 = \alpha \frac{a_1}{a_0} |a_0|^{\alpha-1}.
  \label{eq:45}
\end{equation}
These again yield a self-gravitating wave equation where the static background affects the propagation.

To examine the question of stability we can return to the Lagrangian
formulation of the equations of motion, Eq.~\ref{eq:17}
and~Eq.~\ref{eq:18}.  We can take the time derivative of Eq.~\ref{eq:17} to
get
\begin{equation}
  \frac{d^2 \rho}{dt^2} + \frac{d\rho}{dt} \nabla \cdot {\bf U} + \rho \frac{d}{dt} \nabla \cdot {\bf U} = 0
\label{eq:46}
\end{equation}
and the divergence of Eq.~\ref{eq:18} to yield
\begin{equation}
  \frac{d }{d t} \nabla \cdot {\bf U} + \nabla^2 \left ( V -  \frac{1}{2 a} \nabla^2 a \right ) = 0
\label{eq:47}
\end{equation}
If we have a perturbation on a static solution we find a simpler equation for
the perturbations in the Lagrangian formulation
\begin{equation}
    \frac{d^2 \rho_1}{dt^2} = \nabla^2 \left ( \frac{a_1}{a_0} V_0 + V_1 -  \frac{1}{2 a_0} \nabla^2 a \right ) .
\label{eq:48}
\end{equation}
We will examine a homologous transformation where
\begin{equation}
  {\bf r} = {\bf r}_0 \left (1 + \epsilon \sin \omega t \right ).
  \label{eq:49}
\end{equation}
From Eq.~(\ref{eq:17}) this gives
\begin{equation}
  \rho = \rho_0\left (1 - 3 \epsilon \sin \omega t \right )~\textrm{and}~
  a = a_0\left (1 - \frac{3}{2} \epsilon \sin \omega t \right ).
  \label{eq:50}
\end{equation}
\begin{widetext}
Of course this pertubation is not a solution of Eq.~\ref{eq:48};
however, we can use it to derive an upper bound on the squared frequency of the
oscillation.  From Eq.~\ref{eq:48} we obtain to order $\epsilon$
\begin{equation}
\int dV 3 \epsilon \omega^2 \sin\omega t a_0^2 < \int dV \left [ a_0^\alpha \left ( 1 - \frac{3}{2} \alpha \epsilon \sin \omega t \right) - \left ( 1 - 4 \epsilon \sin \omega t \right ) 
  \nabla^2    \frac{1}{2 a_0} \nabla^2 a_0 \right ],
  \label{eq:51}
\end{equation}
and we can use the zeroth-order solution to simplify this to yield
\begin{equation}
\int dV  3 \epsilon \omega^2 \sin\omega t a_0^2 < \int dV \left [ |a_0|^\alpha \left ( 1 - \frac{3}{2} \alpha \epsilon \sin \omega t \right) - \left ( 1 - 4 \epsilon \sin \omega t \right )
  |a_0|^\alpha \right ]
\label{eq:52}
\end{equation}
\end{widetext}
and
\begin{equation}
 3  \omega^2 \int dV   a_0^2 < \int dV \left ( \frac{8-3\alpha}{2} \right ) |a_0|^\alpha
 \label{eq:53}
\end{equation}
so
\begin{equation}
  \omega^2 < \left ( \frac{8-3\alpha}{6} \right )\int dV |a_0|^\alpha \left [ \int dV a_0^2 \right]^{-1} = \frac{8-3\alpha}{6} \frac{M}{N}
  \label{eq:54}.
\end{equation}
where $M$ is the gravitational mass of the system and $N$ is the
number of particles. Therefore, $\alpha>8/3$ is a sufficient condition
for $\omega^2<0$ and instability for at least one perturbative mode regardless of the static configuration, as we argued from the homology
transformations in \S~\ref{sec:homology}.

If we examine an initially stationary configuration where ${\bf U}\neq
0$ but $d\rho/dt=0$ so $\nabla \cdot {\bf U}=0$, we find to first
order in the perturbation that the same stability condition applies
when one uses the homologous transformation and the variational principle, so we find that
$\alpha>8/3$ is a sufficient condition for instability in general.

\section{Conclusions}
\label{sec:conclusions}

We examine a natural generalization of the Schrodinger-Poisson equation and develop the theory of the static solutions to this equation that we denote quantum polytropes and their stability.  These solutions obey a natural fourth-order generalization of the Lane-Emden equation, the second order equation for classical polytropes.   Furthermore, as for classical polytropes the question of the stability of the solutions comes down to the exponent of the coupling. In the classical case this is how the pressure depends on density with power-law indices greater than $4/3$ indicating stability.  In the quantum case , it is how the boson field generates the gravitational field that leads to instability with power-law indices greater than $8/3$ indicating instability.   We demonstrate the instability in three ways and the criteria all coincide.  We employ two classical techniques, a homology scaling argument and perturbation analysis, and one quantum technique the observation that the states are not bounded from below for $\alpha>8/3$.  This is a sufficient condition for instability not a necessary one.  In particular the excited states even for $\alpha=2$ are unstable \cite{2002math.ph...8045H}. 

The modified Schrodinger-Poisson presented here allows for richer possibilities for the modeling of dark matter halos and structure formation, and  can naturally emerge as the Newtonian limit from an underlying relativistic field theory.
In particular if $\alpha>8/3$ the dark matter halos may develop a quasi-static core that ultimately collapses to form a cusp like standard cold dark matter \cite{1997ApJ...490..493N} or disperses, providing for especially rich phenomenology. 

This work was supported by the Natural Sciences and Engineering Research Council of Canada.

\bibliographystyle{prsty}
\bibliography{non-linear-schrodinger}

\end{document}